\documentclass[11pt,aps,pra,reprint]{revtex4-1}
\usepackage{amsmath}    
\usepackage{graphicx}   
\usepackage{bm}

\def\bra#1{\langle #1 |}
\def\ket#1{| #1 \rangle}
\def\braket#1#2{\langle #1 | #2 \rangle}

\begin{document}
\title{On the validity of the Born-Oppenheimer approximation in the indirect dissociative
recombination process}
\author{Roman \v{C}ur\'{\i}k} \email{roman.curik@jh-inst.cas.cz}
\affiliation{J. Heyrovsk\'{y} Institute of Physical Chemistry, ASCR,
Dolej\v{s}kova 3, 18223 Prague, Czech Republic}
\author{D\'{a}vid Hvizdo\v{s}}
\affiliation{J. Heyrovsk\'{y} Institute of Physical Chemistry, ASCR,
Dolej\v{s}kova 3, 18223 Prague, Czech Republic}
\affiliation{Institute of Theoretical Physics, Faculty of Mathematics and Physics, Charles University in Prague, V Hole\v{s}vi\v{c}k\'{a}ch 2, 180 00 Prague, Czech Republic}
\author{Chris H.~Greene}
\affiliation{Department of Physics and Astronomy, Purdue University, West Lafayette,
Indiana 47907, USA}
\date{\today}

\begin{abstract}
\vspace{-4mm}
An alternative method is introduced to solve a simple 
two-dimensional models describing vibrational
excitation and dissociation processes during the electron-molecule collisions. The
model works with one electronic and one nuclear degree of freedom. The two-dimensional
\textit{R}-matrix can be constructed simultaneously on the electronic and nuclear surfaces
using all three forms developed previously for electron-atom and electron-molecule
collisions. These are the eigenchannel \textit{R}-matrix form, inversion technique of Nesbet
and Robicheaux, and the Wigner-Eisenbud-type form using expansion over the poles of
the symmetrized Hamiltonian. 
The 2D \textit{R}-matrix method is employed to solve a simple model tailored to describe 
the dissociative recombination and the vibrational excitation 
of H$_2^+$ cation in the singlet ungerade symmetry $^1\Sigma_u$.
These results then serve as a (near-exact) benchmark for the following 
calculation in which the \textit{R}-matrix states are replaced by their Born-Oppenheimer
approximations. The accuracy of this approach and its correction with the first-order nonadiabatic
couplings are discussed.
\end{abstract}

\maketitle

\section{\label{sec-intro}Introduction}

Presently there are two generally accepted mechanisms for the dissociative recombination
(DR) of molecular cations. The \textit{direct} 
mechanism involves crossing of the potential curves
of the target system and of the formed neutral molecule \cite{Bates_DR_PR_1950}. 
For several decades this resonant mechanism influenced the theoretical research and 
molecular systems without the curve crossing were assumed to have small DR rates
that were often just estimated 
\cite{Stancil_LD_LiH_ApJ_1996} in early universe chemistry models.

With the increasing number of experimental data in early 1990s, it became difficult to support
this picture in which the curve crossing is required to drive the DR process.
The first theoretical models by \citet{Guberman_HeH_1994} and by
\citet{Sarpal_Morgan_HeH_1994} made it clear that the \textit{indirect}
mechanism, while not requiring a curve crossing, can be quite effective.
Further theoretical studies 
revealed that even for systems with a curve crossing
the DR rate can be enhanced 
\cite{Schneider_OSW_PRL_2000,Kokoouline_Greene_PRA_2003}
or suppressed
\cite{Giusti-Suzor_Derkits_PRA_1983,Nakashima_Nakamura_JCP_1987}
by orders of magnitude when Rydberg states
trigger the initial capture (indirect mechanism).

The vast majority of calculations treating the indirect mechanism
(as examples see Refs.~
\cite{Schneider_Giusti-Suzor_JPB_1991,Curik_Gianturco_2013,Slava_Greene_PRA_2005,
Tanabe_Takagi_HeH_1998,Jugen_Ross_1997,Mezei_DR_BF_PSST_2016} and the references
therein)
employ the quantum defect theory (QDT) in combination with the (ro)vibrational
frame transformation (FT) theory \cite{Chang_Fano_1972}. The frame transformation
approach exploits the Born-Oppenheimer approximation (BOA) that is assumed to be valid
at small electronic distances. The credibility of the FT theory was often tested
by experiments dealing with the elastic and rovibrationally inelastic collisions
of electrons with molecules. However, it is more difficult to carry out similar
comparison for the dissociative recombination process, because
detection of the neutral fragments is more complicated. Moreover,
the target molecular cations are often warmed after they are ionized, and possess an
unknown rovibrational temperature (or distribution) 
before the recombination process takes place
\cite{Semaniak_Zajfman_HeH_1996}.
General agreement between DR theory and experiment 
has frequently been limited to an
order of magnitude and only rarely have detailed experimental fetures been
reproduced
\cite{Chakrabarti_CH_JPB_2018,Curik_Greene_JCP_2017}.
Recent experimental improvements
\cite{CSR_review_2016}, however, have the potential to put the DR theory based on the BOA
to a quantitative test.

In order to assess the accuracy of the Born-Oppenheimer approximation, the cornerstone
of the vibrational frame-transformation theories
\cite{Chang_Fano_1972,Gao_Greene_PRAR_1990,Gao_Greene_JCP_1989,Greene_Jungen_PRL_1985},
we propose 
a numerically solvable two-dimensional (2D) model for the indirect dissociative
recombination of H$_2$ in the singlet ungerade channels $^1\Sigma_u$. 
This model was recently devised
by \citet{Hvizdos_VHGRMC_PRA_2018} to test the accuracy of the energy independent
frame transformation into a nuclear basis of Siegert pseudostates. That study thus provided
a first numerical estimate for the accuracy of the underlying approximations.
The numerically solvable model was based on the exterior complex scaling (ECS) 
applied to both nuclear and electronic coordinates. The ECS approach 
was originally developed to address the dissociative electron attachment and the
vibrational excitation channels in collisions of electrons with neutral molecules
\cite{Houfek_Rescigno_McCurdy_PRA_2006,Houfek_Rescigno_McCurdy_PRA_2008}. For the
target cations the ECS method still provided accurate and converged results for
most of the collision energies
\cite{Hvizdos_VHGRMC_PRA_2018}
, but at comparatively high computational cost due to necessity of using
extensive long-range electronic grids to confine the countless number of Rydberg states involved in 
the closed-channel resonances.

To overcome these difficulties we propose a 2D \textit{R}-matrix method that numerically solves
the electronic-nuclear problem in a 2D box. The size of this box is determined 
by the range of the interaction
that couples the two degrees of freedom. Outer regions, in which either the electron
moves in a pure Coulomb field (or zero field in case of the neutral targets) or the nuclei move
in a constant potential, are treated analytically. This is done by application of the
multichannel quantum defect theory (MQDT)
\cite{Seaton_RPP_1983,Orange_review}
 which is slightly extended to eliminate closed
channels on both the electronic and nuclear surfaces simultaneously.

Finally, the Born-Oppenheimer approximation of the 2D \textit{R}-matrix is also
tested in order to assess the validity of the BOA for the indirect DR process. The
BOA version of \textit{R}-matrix was originally proposed by \citet{Schneider_BoA_Rmat_JPB_1979}
for electron collisions with diatomic molecules. It was applied to explain the boomerang structures
in elastic and vibrationally inelastic electron-N$_2$
\cite{Schneider_BORM_N2_JPB_1979,Morgan_N2_JPB_1986},
and electron-CO \cite{Morgan_CO_JPB_1991} collisions.
Later the method was revived to treat the vibrational
excitation of molecular cations \cite{Sarpal_TM_VE_HeH_JPB_1991,Rabadan_Tennyson_JPB_1999}
and even the DR channel of HeH$^+$ \cite{Sarpal_TM_DR_HeH_JPB_1994}.

\section{\label{sec-rm2d}2D \textit{R}-matrix}

\subsection{\label{ssec-model}The 2D model}

The 2D Hamiltonian partitioning is adopted from Ref.~\cite{Hvizdos_VHGRMC_PRA_2018} and it
is somewhat different from the original notation of 
\citeauthor{Houfek_Rescigno_McCurdy_PRA_2006} 
\cite{Houfek_Rescigno_McCurdy_PRA_2006,Houfek_Rescigno_McCurdy_PRA_2008}.
The present system, with two different modes of fragmentation, associated with the competing
dissociation and ionization (or detachment) channels, will be described by the
time-independent Schr\"{o}dinger equation
\begin{equation}
\label{eq-Schrodinger-2D}
\left[ H_{\mathrm{n}}(R) + H_{\mathrm{e}}(r) + V(R,r) - E\right] \psi(R,r) = 0 \;,
\end{equation}
where 
\begin{eqnarray}
\label{eq-Hn}
H_{\mathrm{n}}(R) &=& - \frac{1}{2 M}\frac{\partial^2}{\partial R^2} + V_0(R)\;, \\
\label{eq-He}
H_{\mathrm{e}}(r) &=& - \frac{1}{2}\frac{\partial^2}{\partial r^2} + 
\frac{l(l+1)}{2 r^2} - \frac{Z}{r}\;.
\end{eqnarray}
The potential curve $V_0(R)$ describes the vibrational motion of the target molecule.
The interaction potential $V(R,r)$ couples the electronic and nuclear degrees of freedom with
the following asymptotic limits:
\begin{alignat}{3}
& V(R,r) = 0 &&\quad \mathrm{for} \quad  && r \geq r_0\;, \\
& V(R,r) = V(R_0,r) && \quad \mathrm{for} \quad && R \geq R_0\;.
\end{alignat}
\begin{figure}[hb]
\includegraphics[width=0.45\textwidth]{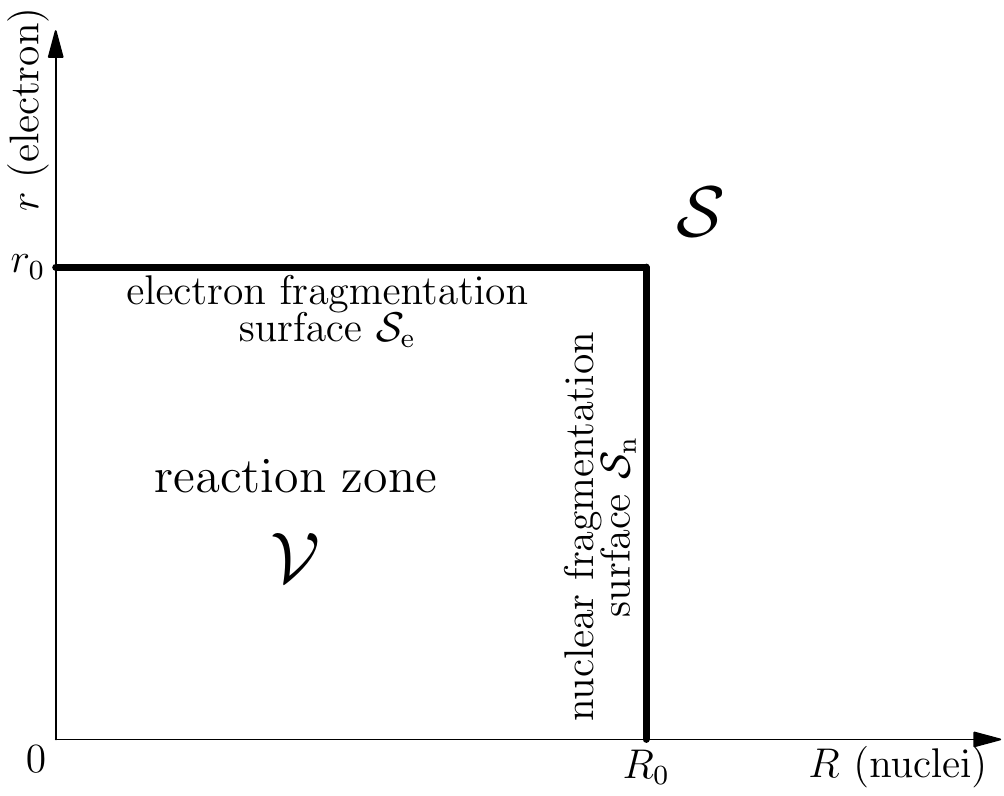}
\caption{\label{fig-rm2df}
Partitioning of the configuration space into inner and outer regions.
}
\end{figure}
In the present study, $V_0$ is chosen to be 
the Morse potential describing the ground state of H$_2^+$, and
$V(R,r)$ is tailored to reproduce approximately the $^1\Sigma_u$ quantum defects
of the neutral hydrogen molecule H$_2$.
The explicit form of the two potentials can be found in Ref.~\cite{Hvizdos_VHGRMC_PRA_2018}.
The charge parameter is $Z=0$ for neutral target molecules, while $Z=1$ for 
target cation. The angular degrees of freedom have already been separated in spherical
coordinates, such that the full solution in the inner region (reaction zone in 
Fig.~\ref{fig-rm2df}) is given by
\begin{equation}
\Psi(\bm{R},\bm{r}) = \frac{1}{R r}\psi(R,r) \Phi(\bm{\Omega})\;,
\end{equation}
where the symbol $\bm{\Omega}$ represents all the angular degrees of freedom. In fact, the
possibility of additional partial waves and electronic symmetries would require
a sum over multiple channels $\Phi$ resulting in coupled set of equations
(\ref{eq-Schrodinger-2D}). In the present study the 2D \textit{R}-matrix method
will be applied to a model describing the dissociative recombination of H$_2^+$ 
in the singlet ungerade symmetry. It is sufficient to assume
\cite{Jungen_Atabek_1977} that only the \textit{p}-wave channel is active for
the electronic coordinate. The nuclear coordinate $R$ is confined 
to \textit{s}-wave scattering and bound vibrational states.

Before we proceed, it is convenient 
\cite{Hamilton_thesis}
to absorb the mass factor $M$ in 
Eq.~(\ref{eq-Hn}) by rescaling of the nuclear coordinate to $X = \sqrt{M}R$.
The change of nuclear variable recasts the full 2D Hamiltonian into a more
symmetric form giving
\begin{equation}
\label{eq-Schrodinger-2DX}
\left[ H_{\mathrm{n}}\left(X\right) + H_{\mathrm{e}}(r) + 
V\left(\frac{X}{\sqrt{M}},r\right) - E\right] \psi(X,r) = 0 \;,
\end{equation}
with
\begin{equation}
H_{\mathrm{n}}\left(X\right) = 
- \frac{1}{2}\frac{\partial^2}{\partial X^2} + V_0\left(\frac{X}{\sqrt{M}}\right)\;.
\end{equation}
For sake of brevity the potentials in the above equations will be denoted simply
as $V(X,r)$ and $V_0(X)$. 

\subsection{\label{ssec-eigR}Eigenchannel \textit{R}-matrix}

Formally, the Schr\"{o}dinger equation (\ref{eq-Schrodinger-2DX}) describes two
interacting distinguishable particles having the same mass of the electron.
In such case the eigenvalues $b(E)$ of the two-particle logderivative operator
$\mathcal{B}(E)$ satisfy the following variational principle
\cite{Szmytkowski_Rmat_PRA_2002}:
\begin{equation}
\label{eq-bstat}
b(E) = 2\; \underset{\psi}{\mathrm{stat}}\left\{ 
\frac{\bra{\psi} \bar{H} -E \ket{\psi}}{\left(\psi|\psi\right)}
\right\}\;,
\end{equation}
where the symmetrized Hamiltonian $\bar{H}$ is defined as
\begin{eqnarray}
\label{eq-bloch1}
\bar{H} &=& H + \frac{1}{2}\left[
\delta(X-X_0)\frac{\partial}{\partial X} +
\delta(r-r_0)\frac{\partial}{\partial r} \right] \;,\\
\label{eq-bloch2}
H &=& H_{\mathrm{n}}(X) + H_{\mathrm{e}}(r) + V(X,r)\;,
\end{eqnarray}
with $X_0 = \sqrt{M}R_0$.
The scalar product $\braket{\psi}{\psi}$ is carried out in the two-particle
volume 
$\mathcal{V} = \langle 0,X_0\rangle\times\langle 0,r_0\rangle$ 
\cite{Szmytkowski_Rmat_PRA_2002} and 
the scalar product denoted by $\left(\psi|\psi\right)$ is carried out on the 
two-particle surface $\mathcal{S}$ enclosing the volume $\mathcal{V}$.
If a surface delta function $\delta(\mathcal{S})$ is defined as
\begin{equation}
\delta(\mathcal{S}) = \delta(X-X_0) + \delta(r-r_0)\;,
\end{equation}
we can simply write
\begin{equation}
\left(\psi|\psi\right) = \bra{\psi} \delta(\mathcal{S}) \ket{\psi}\;.
\end{equation}
The stationary principle (\ref{eq-bstat}) leads to the Schr\"{o}dinger equation
for the eigenvalues $b_\alpha(E)$
\begin{equation}
\label{eq-Schrodinger-b}
2\left(\bar{H} - E \right) \ket{\psi_\alpha} = b_\alpha\, |\psi_\alpha) = 
b_\alpha\, \delta(\mathcal{S}) \ket{\psi_\alpha}\;.
\end{equation}
The logderivative operator $\mathcal{B}(E)$ and the inverse operator
$\mathcal{R}(E) = \mathcal{B}^{-1}(E)$ operate on class of functions
defined on the surface $\mathcal{S}$. In the case these functions are
formed by surface values of $\psi(X,r)$ satisfying 
the 2D Schr\"{o}dinger equation (\ref{eq-Schrodinger-2DX}),
these operators become hermitian
\cite{Szmytkowski_Rmat_PRA_2002}.
The eigenfunctions $\psi_\alpha(X,r)$ are those solutions of 
(\ref{eq-Schrodinger-2DX}) that in addition have common outward normal
logarithmic derivative $b_\alpha$ on the whole surface $\mathcal{S}$. 
They allow a formal
spectral decomposition of $\mathcal{B}(E)$ and $\mathcal{R}(E)$ as
\begin{eqnarray}
\label{eq-Bop-specs}
\mathcal{B} &=& \sum_\alpha |\psi_\alpha)\, b_\alpha\, (\psi_\alpha|\;, \\
\label{eq-Rop-specs}
\mathcal{R} &=& \sum_\alpha |\psi_\alpha)\, b^{-1}_\alpha\, (\psi_\alpha|\;.
\end{eqnarray}
The form of Eq.~(\ref{eq-Rop-specs}) requires a special note related to the
eigensolutions $\psi_\alpha(X,r)$ of Eq.~(\ref{eq-Schrodinger-b}). Since
the surface operator $\delta(\mathcal{S})$ on r.h.s of 
Eq.~(\ref{eq-Schrodinger-b}) has lower rank than the 
symmetrized Hamiltonian on the l.h.s. there
will be, in general, many trivial solutions 
\cite{Orange_review}
with the eigenvalues $b_\alpha = 0$. While these trivial solutions do not
contribute to the spectral form of $\mathcal{B}$ in 
Eq.~(\ref{eq-Bop-specs}) they need to be excluded in the form 
(\ref{eq-Rop-specs}).

The \textit{R}-matrix consists of matrix elements of the $\mathcal{R}$
operator in the basis of functions orthonormal on the surface $\mathcal{S}$.
This basis, also called the fragmentation channel functions, 
can be assembled from two sets of the surface 
solutions. The first set $\phi_{i_\mathrm{e}}(X)$ is defined 
on the electronic surface $\mathcal{S}_\mathrm{e}$
(see Fig.~\ref{fig-rm2df})
\begin{eqnarray}
H_\mathrm{n}(X) \phi_{i_\mathrm{e}}(X) = E_{i_\mathrm{e}} \phi_{i_\mathrm{e}}(X)\;,
\end{eqnarray}
while the second set of surface solutions $\rho_{i_\mathrm{n}}(r)$ 
is defined on the nuclear surface $\mathcal{S}_\mathrm{n}$
\begin{eqnarray}
\left[ H_\mathrm{e}(r) + V(X_0,r) \right] \rho_{i_\mathrm{n}}(r) 
= E_{i_\mathrm{n}} \rho_{i_\mathrm{n}}(r)\;.
\end{eqnarray}
The channels $|i)$ can be then defined on the whole surface $\mathcal{S}$ 
by a union of the two sets ( $i = \{i_\mathrm{e},i_\mathrm{n} \}$ ):
\begin{alignat}{5}
\label{eq-chan1}
&i \in i_\mathrm{e}& &:\; |i) = |\phi_{i_\mathrm{e}})
&\;\mathrm{on}\;& \mathcal{S}_\mathrm{e}
&\;\mathrm{and}\;& |i) = 0 &\;\mathrm{on}\;& \mathcal{S}_\mathrm{n}, \\
\label{eq-chan2}
&i \in i_\mathrm{n}& &:\; |i) = |\rho_{i_\mathrm{n}})
&\;\mathrm{on}\;& \mathcal{S}_\mathrm{n}
&\;\mathrm{and}\;& |i) = 0 &\;\mathrm{on}\;& \mathcal{S}_\mathrm{e}.
\end{alignat}
Continuity of the channel states $|i)$ on the surface $\mathcal{S}$ sets
the boundary conditions for the channel functions:
$\phi_{i_\mathrm{e}}(X_0) = 0$ and $\rho_{i_\mathrm{n}}(r_0) = 0$.
Finally, the \textit{R}-matrix elements in these physically motivated
channels, take simple form in the eigenchannel expression
\begin{equation}
\label{eq-Rmat-eig}
R_{ij} = (i|\mathcal{R}|j) = \sum_\alpha (i|\psi_\alpha)\, 
b^{-1}_\alpha\, (\psi_\alpha|j)\;.
\end{equation}

\subsection{\label{ssec-resR}Resolvent form}

The eigenchannel form (\ref{eq-Rmat-eig}) of the \textit{R}-matrix
is expressed in terms of the eigenvalues and eigenvectors of the
logderivative surface operator $\mathcal{B}$. The resolvent form
was given for one-particle surface by Nesbet 
\cite{Nesbet_book_1980} and by Robicheaux \cite{Robicheaux_1991}.
Its generalization for the present two-particle surface is straightforward.
It is mathematically less awkward here to introduce a 2D basis set
$y_k(R,r)$ in the volume $\mathcal{V}$. The basis set allows
us to express the \textit{R}-matrix (\ref{eq-Rmat-eig}) by an inversion
defined in the volume
\begin{equation}
\label{eq-Rmat-res}
R_{ij} = \frac{1}{2} \sum_{k,l} (i|y_k) (\Gamma^{-1})_{kl} (y_l|j)\;,
\end{equation}
where
\begin{equation}
\Gamma_{kl}(E) = \bra{y_k} (\bar{H}-E) \ket{y_l}\;.
\end{equation}
Both expressions (\ref{eq-Rmat-res}) and
(\ref{eq-Rmat-eig}) are variational forms of the \textit{R}-matrix. However, the
resolvent form (\ref{eq-Rmat-res}) is somewhat easier to 
implement because it requires only a straightforward inversion
(or in practice, an inhomoheneous linear system solution)
of the $(\bar{H}-E)$ term expressed in the 2D basis.
Evaluation through the eigenchannel expression (\ref{eq-Rmat-eig}), on the other
hand, requires a solution of the generalized eigenvalue problem with a singular
matrix on the r.h.s. of Eq.~(\ref{eq-Schrodinger-b}) complemented by a removal of the
trivial solutions. This removal procedure may become problematic if the eigenvalue
$b_\alpha$ of a nontrivial solution $\psi_\alpha$ approaches zero value.

Both \textit{R}-matrix forms, (\ref{eq-Rmat-res}), and (\ref{eq-Rmat-eig})
become computationally demanding in situations in which the \textit{R}-matrix
needs to be evaluated repeatedly for many total energies $E$. In such cases,
the Wigner-Eisenbud expansion over the poles of 
$\bar{H}$
is more efficient.

\subsection{\label{ssec-polR}Wigner-Eisenbud expansion}

Equivalence of the Wigner-Eisenbud 
\cite{Wigner_Eisenbud_1947}
expansion of the \textit{R}-matrix
and its resolvent form (\ref{eq-Rmat-res}) was de\-mon\-stra\-ted, for one-par\-tic\-le
sur\-fa\-ce, by Ro\-bi\-cheaux \cite{Robicheaux_1991}. Pre\-sent two-particle case follows
closely the same idea of a spectral decomposition of the operator
\begin{equation}
\label{eq-Hbar-spect}
\left(\bar{H}-E\right)^{-1} = \sum_p
\frac{\ket{\psi_p}\bra{\psi_p }}{E_p - E}\;,
\end{equation}
where the eigenstates $\ket{\psi_p}$ and eigenvalues $E_p$
are defined by 
\begin{equation}
\label{eq-Hbar-eig}
\bar{H} \ket{\psi_p} = 
E_p \ket{\psi_p}\;.
\end{equation}
Combination of Eqs.~(\ref{eq-Rmat-res}) and (\ref{eq-Hbar-spect}) leads to the
Wigner-Eisenbud expansion of the \textit{R}-matrix, which can be written in a
form independent of the basis set, as
\begin{equation}
\label{eq-Rmat-WE}
R_{ij} = \frac{1}{2}\sum_p
\frac{(i|\psi_p)\,(\psi_p|j)}
{E_p - E}\;.
\end{equation}
It is important to emphasize that the eigenstates $\ket{\psi_\alpha}$ of 
Eqs.~(\ref{eq-Schrodinger-b}--\ref{eq-Rop-specs}) and 
$\ket{\psi_p}$ of Eqs.~(\ref{eq-Hbar-spect}--\ref{eq-Rmat-WE})
are different. While $\ket{\psi_\alpha}$ solve Schr\"{o}dinger equation
(\ref{eq-Schrodinger-2DX}) for a given total energy $E$, the states 
$\ket{\psi_p}$ satisfy this equation only for $E = E_p$.
Even for these discrete energies the two sets of eigenstates differ as they
possess different boundary conditions on the surface $\mathcal{S}$.

\subsection{\label{ssec-Rboa}Adiabatic expansion}

All three of the \textit{R}-matrix forms presented in 
Sections~\ref{ssec-eigR}, \ref{ssec-resR}, and \ref{ssec-polR} provide
information on the surface logarithmic derivative of the exact 2D model 
solution.
In order to assess validity of the Born-Oppenheimer
approximation, we expand the exact 2D eigenstates
$\psi_p(X,r)$ in Eqs.~(\ref{eq-Hbar-eig}) and (\ref{eq-Rmat-WE}) 
into the fixed-nuclei solutions $\psi_k(r;X)$ as follows,
\begin{equation}
\label{eq-adiabexp}
\psi_p(X,r) = \sum_{k'} \psi_{k'}(r;X) \phi_{k'p}(X)\;,
\end{equation}
where the electronic solutions diagonalize the fixed-nuclei Hamiltonian,
\begin{equation}
\label{eq-Ek}
\left[ \bar{H}_{\mathrm{e}}(r) + V(X,r) \right] \psi_k(r;X) 
= \bar{E}_k(X) \psi_k(r;X)\;,
\end{equation}
and the nuclear functions $\phi_{k'p}(X)$ result from the coupled set of
nuclear Schr\"{o}dinger equations
\begin{multline}
\label{eq-nuc-coupl}
\left[ \bar{H}_{\mathrm{n}} + \bar{E}_k(X) - E_p \right] \phi_{k p}(X) = \\
-\frac{1}{2} \sum_{k'} \left[ V^{(1)}_{k k'}(X) + V^{(2)}_{k k'}(X) \right] 
\phi_{k'p}(X). 
\end{multline}
The first-order nonadiabatic coupling operator for the symmetrized 
Hamiltonian $\bar{H}$ can be written as
\begin{equation}
\label{eq-Vc1}
V^{(1)}_{k k'}(X) = \ket{\frac{d}{d X}} \braket{\psi_k}{\psi'_{k'}}_r
+ \braket{\psi_k'}{\psi_{k'}}_r \bra{\frac{d}{d X}},
\end{equation}
where $\psi'_k = \partial \psi_k(r;X) / \partial X$ and the scalar
product $\braket{.}{.}_r$ is carried out only on the electronic coordinate $r$.
The first-order nuclear derivative in the first term acts "to the left", e.g. when
matrix elements in the nuclear basis are evaluated.
The second-order nonadiabatic terms have the form of local potentials
\begin{equation}
\label{eq-Vc2}
V^{(2)}_{k k'}(X) = \braket{\psi'_k}{\psi'_{k'}}_r\;.
\end{equation}
Finally, the symmetrized nuclear and electronic Hamiltonians $\bar{H}_{\mathrm{n}}$
and $\bar{H}_{\mathrm{e}}$ are obtained by splitting the Bloch operator on
the r.h.s. of Eq.~(\ref{eq-bloch1}) into respective nuclear and electronic parts,
i.e.
\begin{eqnarray}
\bar{H}_{\mathrm{n}}(X) &=& H_{\mathrm{n}}(X) + 
\frac{1}{2} \delta(X-X_0)\frac{\partial}{\partial X}\;,\\
\bar{H}_{\mathrm{e}}(r) &=& H_{\mathrm{e}}(r) +
\frac{1}{2} \delta(r-r_0)\frac{\partial}{\partial r} \;.
\end{eqnarray}

The Born-Oppenheimer approximation neglects the nonadiabatic couplings
$V^{(1)}_{k k'}$ and $V^{(2)}_{k k'}$. In this case the set of equations
(\ref{eq-nuc-coupl}) decouple and only one term survives in expansion (\ref{eq-adiabexp})
\begin{equation}
\psi^{\mathrm{BO}}_p(X,r) = \psi_k(r;X) \phi_{kn}(X)\;,
\end{equation}
where $p\equiv\{k,n\}$ represents a combined index of electronic states (indexed by $k$)
and nuclear states (indexed by $n$).
The \textit{R}-matrix 
\begin{equation}
\label{eq-Rmat-Schneider}
R^{\mathrm{BO}}_{ij} = \frac{1}{2}\sum_p
\frac{(i|\psi^\mathrm{BO}_p)\,(\psi^\mathrm{BO}_p|j)}
{E_p - E}\;,
\end{equation}
based on the Born-Oppenheimer states was introduced previously for the diatomic
molecules by
\citet{Schneider_BoA_Rmat_JPB_1979}.

In the final note of this section we would like to discuss the radius
$r_0$ of the electronic box beyond which we assume $V(X,r) = 0$.
On one side we would prefer to have $r_0$ as small as possible to 
improve accuracy of the Born-Oppenheimer 
\textit{R}-matrix~(\ref{eq-Rmat-Schneider}). On another side, for very small
$r_0$ values, the electronic channels $\rho_{i_\mathrm{n}}(r)$ may not
fit into the nuclear fragmentation surface
$\mathcal{S}_\mathrm{n}$ (see Fig.~\ref{fig-rm2df}). This issue was previously
discussed by \citet{Jungen_PRL_1984} where the
author introduced radius $r_2$ as a "distance where all relevant bound Rydberg
components have fallen exponentially to a negligibly small value". 
In the present
study first two electronic states are open already at zero collision energy and
they barely fit into the box size of $r_2 = 20$ bohr,
while $V(X,r)$ can be considered zero beyond $r_0$ = 6-7 bohr.
Therefore, these two contradicting requirements lead to a question, namely whether an
\textit{R}-matrix determined in a small 2D box confined by $r_0$ can be losslessly
propagated onto a surface of a larger 2D box confined by $r_2$, while the nuclear
box size $X_0$ does not change. Such a technique for one-dimensional
\textit{R}-matrix propagation was developed by
\citet{Baluja_BM_CPC_1982}. The generalization
of this procedure for propagation of the 2D \textit{R}-matrix, needed for the
present problem, can be found in the Appendix.

\subsection{\label{ssec-outer}Outer region (cation case)}

In this study we do not attempt to solve a dissociative scattering
problem in which $R\rightarrow\infty$ and $r\rightarrow\infty$ simultaneously.
Instead, we assume that at least one of the coordinates $R$ or $r$ is confined to
the range
$R \leq R_0$ or $r \leq r_0$, respectively. This restriction is also reflected
in the choice of the surface channels 
(\ref{eq-chan1}), (\ref{eq-chan2})
that always vanish at the point where $\mathcal{S}_{\mathrm{n}}$ and
$\mathcal{S}_{\mathrm{e}}$ meets.
In the following we consider $N_\mathrm{e}$ channel functions
$\phi_{i_\mathrm{e}}(X)$ on the electronic
surface $\mathcal{S}_\mathrm{e}$ and $N_\mathrm{n}$ channel functions
$\rho_{i_\mathrm{n}}(r)$ on the nuclear
surface $\mathcal{S}_\mathrm{n}$. Therefore, the total number of the surface
channels defined by Eqs.~(\ref{eq-chan1}), (\ref{eq-chan2}) 
is $N_\mathrm{e}+N_\mathrm{n}$. 

Because the Hamiltonian (\ref{eq-bloch2}) becomes separable on the surface 
$\mathcal{S}$, the solutions in the outer regions are made as a sum of products of
the channel functions and of the asymptotic solutions. The two independent
electronic solutions describing the electronic fragmentation in the Coulomb 
field will be denoted as $f(r)$ and $g(r)$. Analytic properties of the
Coulomb functions $(f,g)$ are
detailed completely by \citet{Seaton_RPP_1983} 
(who calls them $(s,-c)$ functions). For positive channel energies
$\epsilon_{i_\mathrm{e}} = E - E_{i_\mathrm{e}}$ they have an asymptotic limit 
of harmonic functions with the energy normalization
\begin{alignat}{2}
f_{i_\mathrm{e}}(r) &\rightarrow& &(2/\pi k_{i_\mathrm{e}})^{1/2}
\sin(k_{i_\mathrm{e}}r + (1/k_{i_\mathrm{e}})\ln r + \eta )\;, \\
g_{i_\mathrm{e}}(r) &\rightarrow& -&(2/\pi k_{i_\mathrm{e}})^{1/2}
\cos(k_{i_\mathrm{e}}r + (1/k_{i_\mathrm{e}})\ln r + \eta ) \;,
\end{alignat}
where $\eta(k_{i_\mathrm{e}},l)$ is a long-range phase shift \cite{Orange_review} and
the channel momenta are defined by $k_{i_\mathrm{e}}^2/2 = \epsilon_{i_\mathrm{e}}$.
For negative channel energies both functions $f_{i_\mathrm{e}}(r)$ and
$g_{i_\mathrm{e}}(r)$ contain exponentially growing and decaying parts 
\cite{Seaton_RPP_1983,Orange_review}.

For the asymptotic region beyond the nuclear fragmentation surface 
$\mathcal{S}_{\mathrm{n}}$ we use zero-field \textit{s}-wave 
radial functions
\begin{alignat}{2}
\label{eq-FG0-pos1}
F^0_{i_\mathrm{n}}(X) &\rightarrow& &(2/\pi)^{1/2} K_{i_\mathrm{n}}^{-1}
\sin(K_{i_\mathrm{n}} X) \;, \\
\label{eq-FG0-pos2}
G^0_{i_\mathrm{n}}(X) &\rightarrow& - &(2/\pi)^{1/2} 
\cos(K_{i_\mathrm{n}} X) \;, 
\end{alignat}
for positive $\epsilon_{i_\mathrm{n}} = E - E_{i_\mathrm{n}}$ and
\begin{alignat}{2}
\label{eq-FG0-neg1}
F^0_{i_\mathrm{n}}(X) &\rightarrow& &(1/2\pi)^{1/2} \kappa_{i_\mathrm{n}}^{-1}
\left(e^{\kappa_{i_\mathrm{n}}X} - e^{-\kappa_{i_\mathrm{n}}X} \right)\;,\\
\label{eq-FG0-neg2}
G^0_{i_\mathrm{n}}(X) &\rightarrow& -&(1/2\pi)^{1/2} 
\left(e^{\kappa_{i_\mathrm{n}}X} + e^{-\kappa_{i_\mathrm{n}}X} \right)\;,
\end{alignat}
for the negative $\epsilon_{i_\mathrm{n}} = -\kappa^2_{i_\mathrm{n}}/2$.
The asymptotic functions $F^0_i$ and $G^0_i$ are not energy normalized, however, they can
be smoothly continued through the zero channel energy. Eventually, the wave function for
fragmentation regions will
be written in terms of the energy-normalized nuclear asymptotic functions, but 
but the respective transformation will be postponed to the later stage of the
present treatment of the outer region.
If the diagonal matrices for the asymptotic functions evaluated on the whole surface $\mathcal{S}$
are constructed as follows,
\begin{eqnarray}
\!\!\!\!\underline{\mathcal{F}}\!&=&\!\mathrm{diag}\left[
f_1(r_0),...\,,\!f_{N_\mathrm{e}}(r_0),F^0_1(X_0),...\,,\!F^0_{N_\mathrm{n}}(X_0)\right],\\
\!\!\!\!\underline{\mathcal{G}}\!&=&\! \mathrm{diag}\left[
g_1(r_0),...\,,\!g_{N_\mathrm{e}}(r_0),G^0_1(X_0),...\,,\!G^0_{N_\mathrm{n}}(X_0)
\right],
\end{eqnarray}
the short-range \textit{K}-matrix $\underline{K}$ describing the wave function 
in both fragmentation regions can be expressed by a familiar transformation
\begin{equation}
\label{eq-R2K}
\underline{K} = \left( \underline{\mathcal{F}} -
\underline{\mathcal{F}}' \underline{R} \right)
\left( \underline{\mathcal{G}} -
\underline{\mathcal{G}}' \underline{R} \right)^{-1}\;.
\end{equation}

Because no asymptotic boundary conditions have been enforced up to this point,
the independent solutions in both fragmentation regions
\begin{multline}
\label{eq-psiK}
\psi_{i'}(X,r) = \sum_{i_\mathrm{e}=1}^{N_\mathrm{e}} \phi_{i_\mathrm{e}}(X)
\left[ f_{i_\mathrm{e}}(r) \delta_{i_\mathrm{e} i'} -
g_{i_\mathrm{e}}(r) K_{i_\mathrm{e} i'} \right]\\
+
\sum_{i_\mathrm{n}=1}^{N_\mathrm{n}} \rho_{i_\mathrm{n}}(r)
\left[ F^0_{i_\mathrm{n}}(X) \delta_{i_\mathrm{n} i'} -
G^0_{i_\mathrm{n}}(X) K_{i_\mathrm{n} i'} \right],
\end{multline}
contain exponentially growing components for $r\rightarrow\infty$
or $X\rightarrow\infty$.
Within the MQDT formalism the exponentially growing components of 
$\psi_{i'}$ are cancelled by a proper linear combinations of these
functions. There are two MQDT techniques available to carry out this 
elimination of closed channels in the case of one-particle fragmentation. 
The first technique 
\cite{Orange_review,Greene_Jungen_AAMP_1985,Nakashima_Nakamura_JCP_1987}
works with the short-range \textit{K}-matrix (\ref{eq-psiK}) 
or \textit{S}-matrix expressed in asymptotic channels. This procedure
leads to a well-known inversion
formula for the physical \textit{K}- or \textit{S}-matrix that are defined
in the space of open channels.
The second technique 
\cite{Greene_Jungen_AAMP_1985,Lu_PRA_1971,Gao_Greene_JCP_1989}
is based on the eigenchannel representation
of the wave function in one-particle asymptotic region and it exploits
the fact that the asymptotic phases of the eigenchannel solutions
are equal in all the channels. In the present two-particle fragmentation
procedure we adopt the latter, the eigenchannel approach.

The eigenchannel solutions 
\begin{multline}
\label{eq-psi-eig}
\psi_{\gamma}(X,r)\!=\!\!\!
\sum_{i_\mathrm{e}=1}^{N_\mathrm{e}}\!\! \phi_{i_\mathrm{e}}(X)
U_{i_\mathrm{e}\gamma}\!
\left[ f_{i_\mathrm{e}}(r) \cos\pi\tau_\gamma \!-\!
g_{i_\mathrm{e}}(r) \sin\pi\tau_\gamma \right]\\
+ 
\sum_{i_\mathrm{n}=1}^{N_\mathrm{n}} \rho_{i_\mathrm{n}}(r) U_{i_\mathrm{n}\gamma}
\left[ F^0_{i_\mathrm{n}}(X) \cos\pi\tau_\gamma -
G^0_{i_\mathrm{n}}(X) \sin\pi\tau_\gamma \right],
\end{multline}
have common eigenphase in all the electronic and nuclear fragmentation channels
\cite{Jugen_Ross_1997}.
Here $\tan\pi\tau_\gamma$ and $U_{i\gamma}$ are the eigenvalues and the orthonormal
eigenvectors of $\underline{K}$ (\ref{eq-R2K}), respectively.
Physical boundary conditions at $r\rightarrow\infty$ and $X\rightarrow\infty$
can be enforced by proper linear combination of the eigensolutions, i.e.
\begin{equation}
\label{eq-psi-elim}
\psi(X,r) = \sum_\gamma \psi_\gamma A_\gamma \;.
\end{equation}
The coefficients $A_\gamma$ must be found such that the wave function 
$\psi(X,r)$ decays exponentially in each closed channel on the electronic
fragmentation surface $\mathcal{S}_{\mathrm{e}}$ ($i_{\mathrm{e}} 
\in Q_{\mathrm{e}}$) and also in every closed channel on the nuclear 
fragmentation surface $\mathcal{S}_{\mathrm{n}}$, i.e. for 
$i_{\mathrm{n}} \in Q_{\mathrm{n}}$.
Secondly, the wave function (\ref{eq-psi-elim}) must approach the
physical eigenchannel solution which requires a common physical
eigenphase shift $\delta$ in each of the $N^\mathrm{o}_\mathrm{e}$ 
open electronic channels ($i_{\mathrm{e}}
\in P_{\mathrm{e}}$) and also in every of the $N^\mathrm{o}_\mathrm{n}$
open nuclear channels, i.e. for all
$i_{\mathrm{n}} \in P_{\mathrm{n}}$.
By combining the Coulomb and free-field procedures described in detail
in Refs. \cite{Greene_Jungen_AAMP_1985} and 
\cite{Gao_Greene_JCP_1989} it can be shown that there can be no more than
$N^\mathrm{o}_\mathrm{e} + N^\mathrm{o}_\mathrm{n}$ such coefficient
sets $A_\gamma$ that lead to the $\psi(X,r)$ satisfying all these conditions.
These different coefficients sets will be distinguished by a second index
that gives a matrix $A_{\gamma\rho}$. Moreover, the column vectors of 
$\underline{A}$ are eigenvectors of a singular generalized eigenvalue problem
\begin{equation}
\underline{\Gamma}\, \underline{A} = \underline{\Lambda}\, \underline{A}
\tan \underline{\delta},
\end{equation}
with
\begin{equation}
\label{eq-chanelim-G}
\Gamma_{i\gamma} = \left\{
\begin{array}{ll}
U_{i\gamma} \sin(\beta_i + \pi\tau_\gamma), & i \in Q_{\mathrm{e}} \\
U_{i\gamma} \left( \kappa_i^{-1} \cos\pi\tau_\gamma + \sin\pi\tau_\gamma \right), &
i \in Q_{\mathrm{n}} \\
\\
U_{i\gamma} \sin\pi\tau_\gamma, & i \in P_{\mathrm{e}} \\
U_{i\gamma} K_i^{1/2} \sin\pi\tau_\gamma, & i \in P_{\mathrm{n}} 
\end{array}
\right. ,
\end{equation}
and
\begin{equation}
\label{eq-chanelim-L}
\Lambda_{i\gamma} = \left\{
\begin{array}{ll}
0, & i \in Q_{\mathrm{e}} \\
0, & i \in Q_{\mathrm{n}} \\
\\
U_{i\gamma} \cos\pi\tau_\gamma, & i \in P_{\mathrm{e}} \\
U_{i\gamma} K_i^{-1/2} \cos\pi\tau_\gamma, & i \in P_{\mathrm{n}} 
\end{array}
\right. .
\end{equation}
The MQDT symbol $\beta_i$ denotes effective Rydberg quantum numbers with respect to the
closed-channel thresholds $E_i$,
\begin{equation}
\beta_i = \frac{\pi}{\sqrt{2(E_i - E)}} \;.
\end{equation}

The $K^{1/2}_i$ terms in Eqs.~(\ref{eq-chanelim-G}) and (\ref{eq-chanelim-L}) allow
to write the open eigenchannel solutions in terms of the energy-normalized asymptotic
functions in both fragmentation regions as
\begin{multline}
\label{eq-psi-eigo}
\psi_{\rho}(X,r) = 
\sum_{i \in P_{\mathrm{e}}} \phi_i(X)
T_{i\rho}
\left[ f_i(r) \cos\pi\delta_\rho -
g_i(r) \sin\pi\delta_\rho \right]\\
+ 
\sum_{i \in P_{\mathrm{n}}} \rho_i(r) T_{i\rho}
\left[ F_i(X) \cos\pi\delta_\rho - 
G_i(X) \sin\pi\delta_\rho \right],
\end{multline}
where the energy-normalized nuclear functions are related to the analytic functions
(\ref{eq-FG0-pos1})-(\ref{eq-FG0-neg2}) by relations 
\cite{Gao_Greene_JCP_1989}
$F_i(X) = K^{1/2}_i F^0_i(X)$ and
$G_i(X) = K^{-1/2}_i G^0_i(X)$. The transformation matrix of open eigenchannels
\begin{equation}
T_{i\rho} = \sum_\gamma A_{\gamma\rho} \left(
\Lambda_{i\gamma} \cos\pi\delta_\rho + \Gamma_{i\gamma} \sin\pi\delta_\rho \right),
\end{equation}
is orthogonal and it can be made orthonormal by choosing an appropriate
normalization of the
eigenvectors $\underline{A}$.

The physical scattering matrix $\underline{S}$ has dimension of 
$N^\mathrm{o}_{\mathrm{e}} + N^\mathrm{o}_{\mathrm{n}}$ and it can be obtained from
the open eigenchannels as
\begin{equation}
S_{i j} = \sum_\rho T_{i\rho} e^{2 i \delta_\rho} T_{j\rho},\quad
i,j \in P_{\mathrm{e}} \cup P_{\mathrm{n}},
\end{equation}
leading to vibrationally inelastic and dissociative recombination integral
cross sections
\begin{alignat}{2}
\sigma^{\mathrm{VE}}_{i \leftarrow j} &= \frac{\pi}{2\epsilon_j} |S_{i j} - \delta_{i j}|^2,
\quad &i,j \in P_{\mathrm{e}}, \\
\sigma^{\mathrm{DR}}_{i \leftarrow j} &= \frac{\pi}{2\epsilon_j} |S_{i j}|^2, 
\quad &i \in P_{\mathrm{n}}, j \in P_{\mathrm{e}}.
\end{alignat}

\section{\label{sec-res}Results and discussion}

The exact Hamiltonian (\ref{eq-Hbar-eig}) together with 
the BOA Hamiltonians (\ref{eq-Ek}) and 
(\ref{eq-nuc-coupl}) have been diagonalized in the 2D box confined by $R_0$ = 15 bohr radii
and $r_2$ = 50 bohr radii. Large electronic box is chosen to properly represent all
the electronic states up to $n=4$ into which the nuclei dissociate for the
examined collision energy 0--2~eV.

The 2D basis was represented as a product of one-dimensional
B-splines \cite{Bachau_CDHM_RPP_2001}. Electronic functions are fairly smooth over
all the examined collision energy range and it was sufficient to
involve about 50-60 B-splines for the electronic coordinate. Momenta are larger
in the nuclear coordinate and therefore we needed about 80 B-splines to converge
the DR into $n=2$ state. However, the nuclei have more than 37~eV of kinetic energy
when dissociating into (unphysical) $n=1$ state and the convergence for this channel required
about 200 of B-spline functions.

\begin{figure}[tbh]
\includegraphics[width=0.46\textwidth]{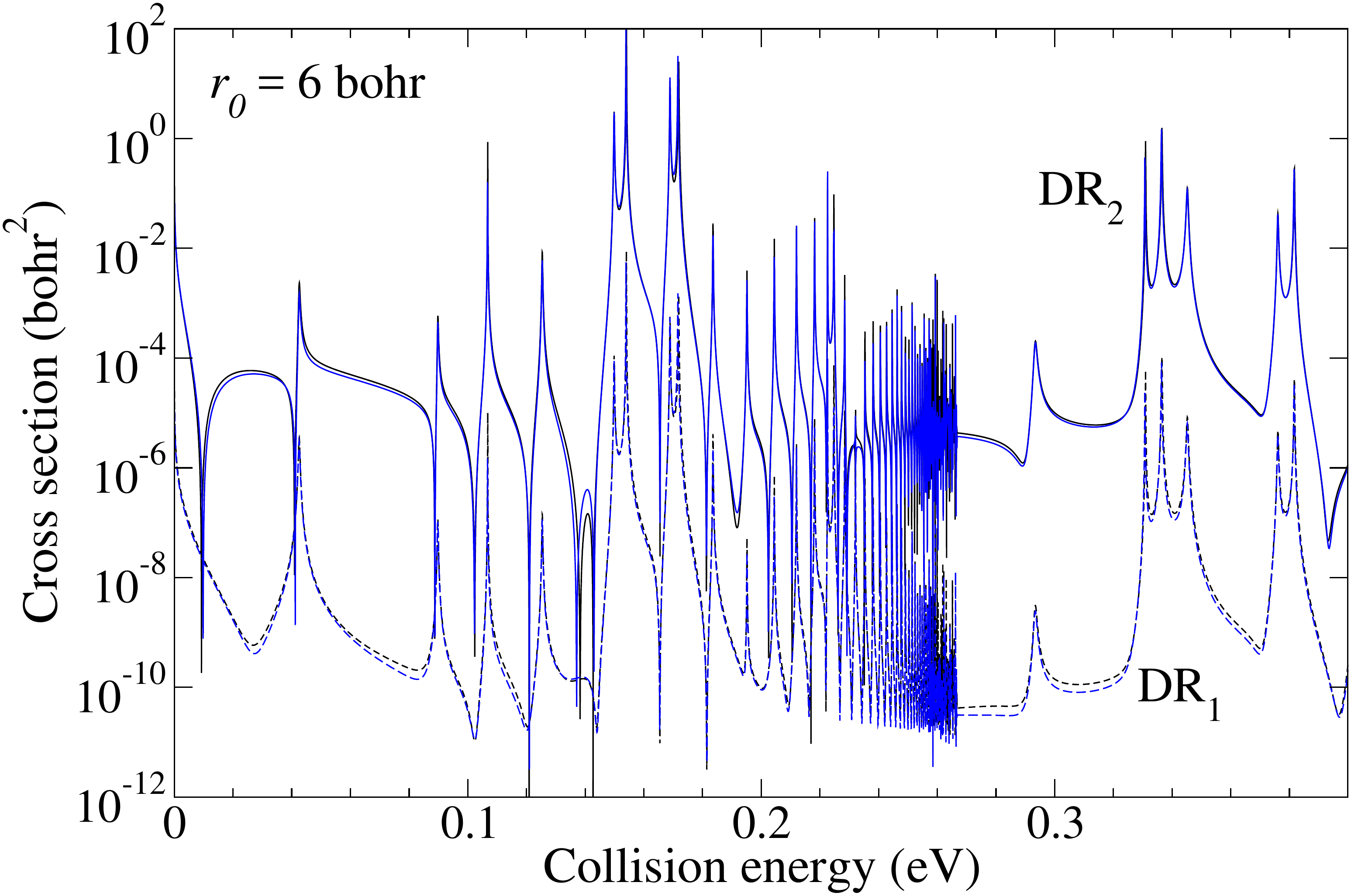}
\caption{\label{fig-boa_r06}
DR cross sections into final $n=1$ (broken curves) and $n=2$ states (full lines).
Black lines are exact results, while the blue lines show calculations from
the Born-Oppenheimer \textit{R}-matrix and $r_0$ = 6 bohr.
}
\end{figure}
\begin{figure}[tbh]
\includegraphics[width=0.46\textwidth]{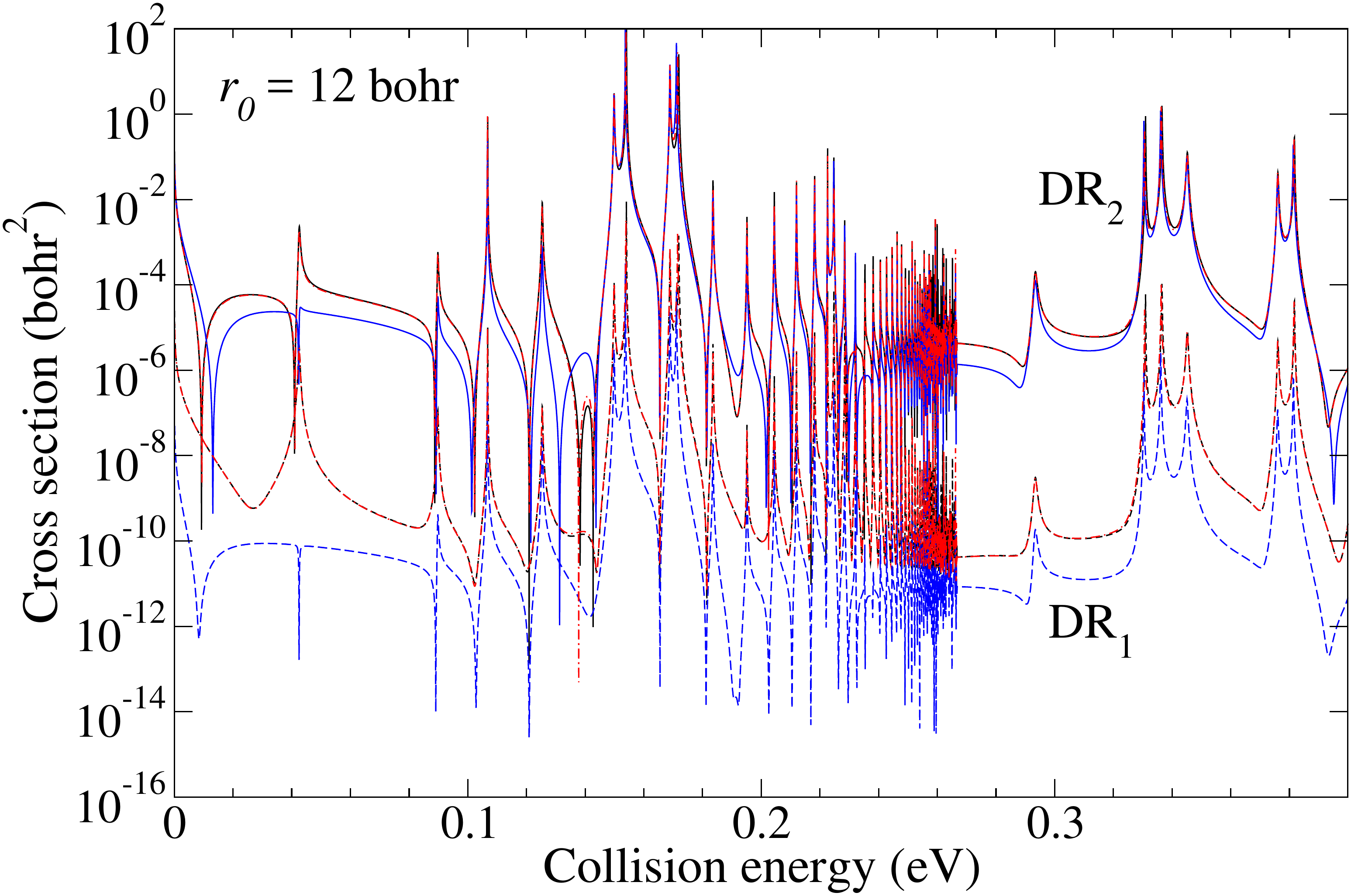}
\caption{\label{fig-boa_r12}
DR cross sections into final $n=1$ (broken curves) and $n=2$ states (full lines).
Black lines are exact results, while the red lines show calculations from
the Born-Oppenheimer \textit{R}-matrix and $r_0$ = 12 bohr. Red dot-dashed curves
represent results with first order non-adiabatic couplings included.
}
\end{figure}
\begin{figure}[tbh]
\includegraphics[width=0.46\textwidth]{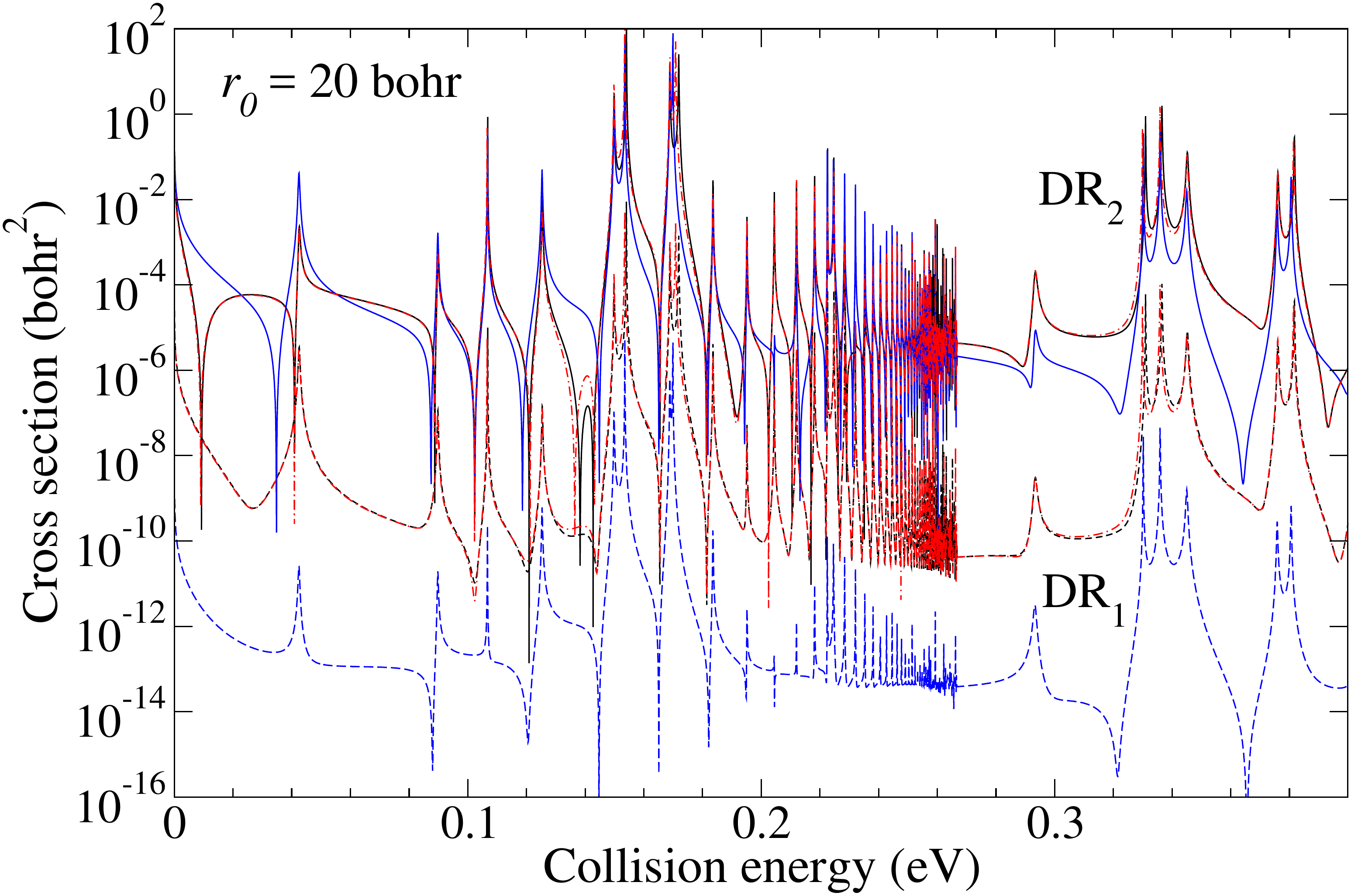}
\caption{\label{fig-boa_r20}
Same as in Fig.~\ref{fig-boa_r12} with the
radius $r_0$ = 20 bohr.
}
\end{figure}
All the three exact forms of the 2D \textit{R}-matrix 
(\ref{eq-Rmat-eig}), (\ref{eq-Rmat-res}), and (\ref{eq-Rmat-WE})
yielded the same numerical results. However, for the repeated evaluation
of the \textit{R}-matrix on a dense energy grid, the Wigner-Eisenbud
form (\ref{eq-Rmat-WE}) is the most convenient one.

In the second set of calculations the exact \textit{R}-matrix
was replaced by its Born-Oppenheimer approximation (\ref{eq-Rmat-Schneider}).
Since the validity of BOA strongly depends on the electronic box size we need to propagate
the \textit{R}-matrix determined at small $r_0$ to $r_2$ = 50 bohr to satisfy the conditions
at which the exact results were obtained. For this we employed the technique devised in
the Appendix.

In the third set of calculations we attempt to correct the Born-Oppenheimer results
by involvement of the first-order nonadiabatic couplings (\ref{eq-Vc1}) in
Eq.~(\ref{eq-nuc-coupl}). Obviously, inclusion of the both first-order 
and second-order (\ref{eq-Vc2}) couplings reconstructs the
exact results accurately. 

Comparison between the exact results and the BOA results is shown in 
Fig.~\ref{fig-boa_r06}. The \textit{R}-matrix radius $r_0$ = 6 bohr is the lowest
possible value that confines the interaction $V(R,r)$ in Eq.~(\ref{eq-Schrodinger-2D})
and thus it represents the best possible conditions for validity of the BOA.
The collision energy range chosen for the demonstration is 0--400 meV. The results
were computed and analyzed up to 2~eV and they all follow the conclusions that
will be demonstrated on this lower energy window. It is clear that for both DR channels that
are open at these energies the BOA \textit{R}-matrix very successfully reconstructs
the exact results. For this case we do not show the first-order corrected results
because they are practically identical with the exact numbers.

This situation changes already for $r_0$ = 12 bohr. The BOA cross sections displayed in
Fig.~\ref{fig-boa_r12} show visible deviations from the exact results. The dominant
$n$ = 2 channel cross section is several times lower than the exact results,
 and the weaker $n$ = 1 channel differs by 
1-2 orders of magnitude. Once the first-order nonadiabatic coupling terms are included,
the results reconstruct the exact numbers accurately. For most of the collision
energies shown they are barely distinguishable in Fig.~\ref{fig-boa_r12}.

\begin{figure}[tbh]
\includegraphics[width=0.46\textwidth]{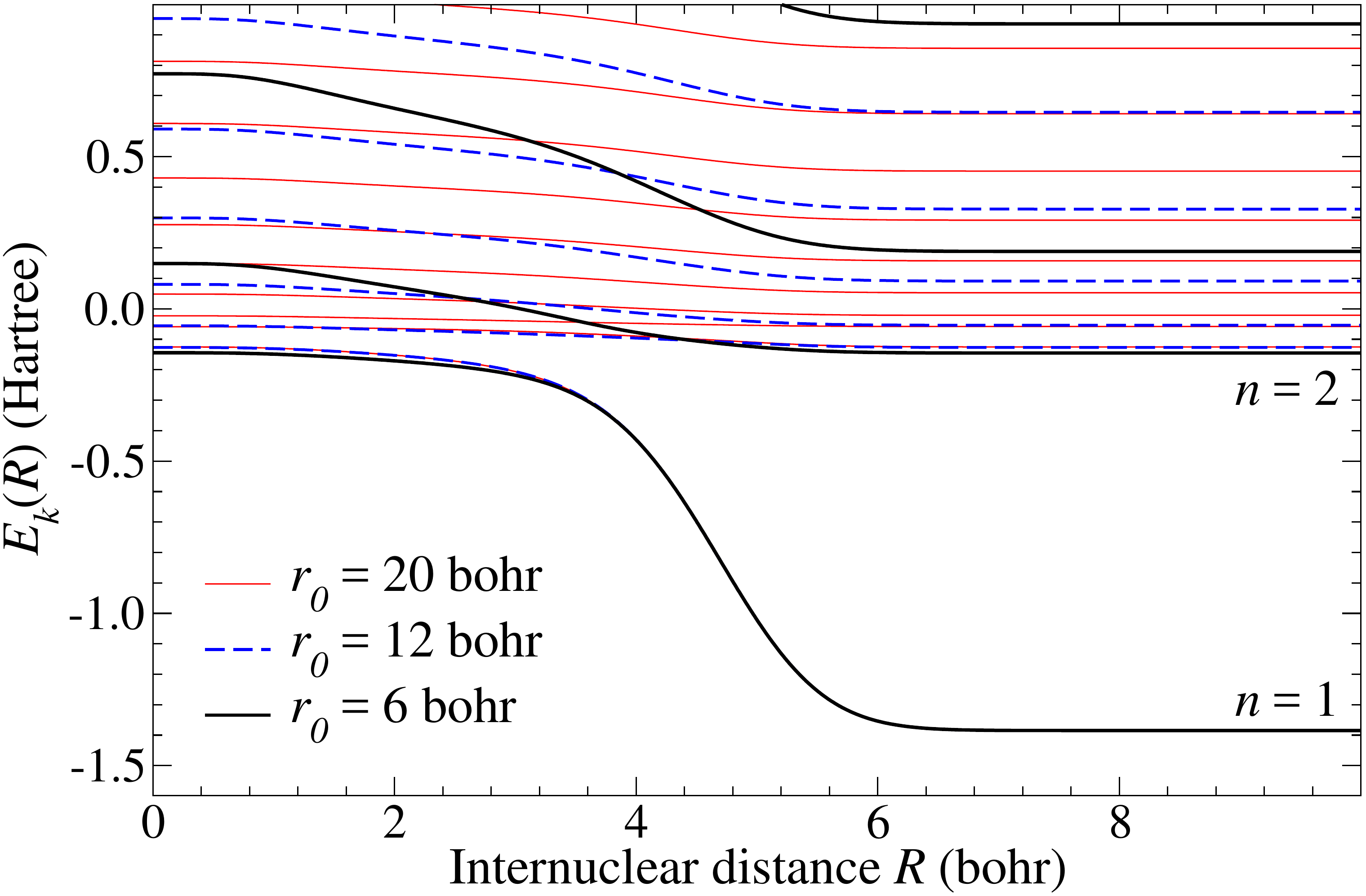}
\caption{\label{fig-Ep}
Fixed-nuclei \textit{R}-matrix poles $\bar{E}_k(R)$ (\ref{eq-Ek}) as a function of the
internuclear distance $R$. Energy curves are shown for three sizes $r_0$ of the 
\textit{R}-matrix box.
}
\end{figure}
\begin{figure}[tbh]
\includegraphics[width=0.46\textwidth]{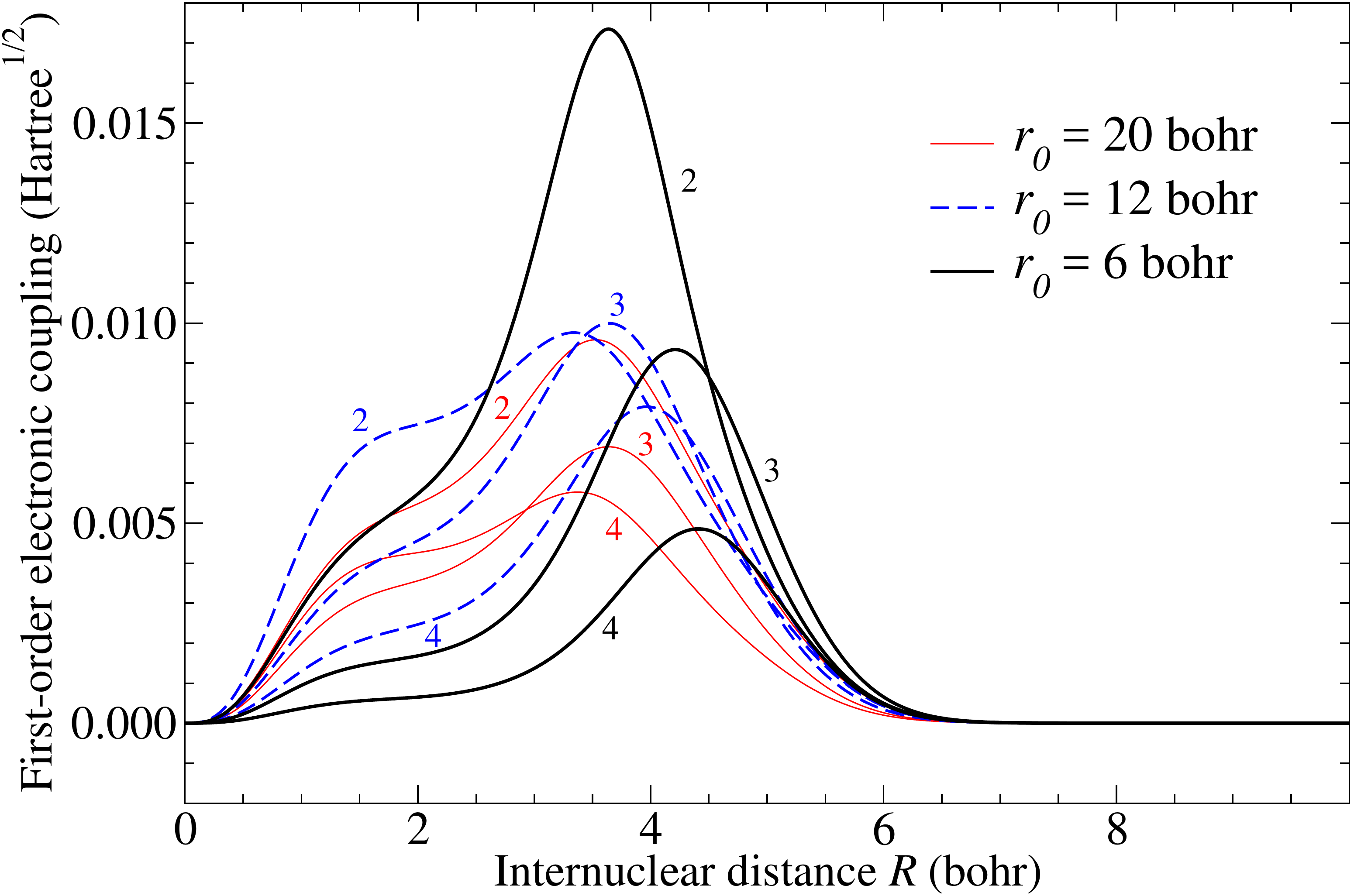}
\caption{\label{fig-Vc1}
First-order electronic coupling terms $\braket{\psi_k'}{\psi_{k'}}_r$ of
Eq.~(\ref{eq-Vc1}), where $k'$ = 1 and $k$ = 2,3,4. Data for three 
\textit{R}-matrix box sizes $r_0$ are displayed.
}
\end{figure}

Validity of the BOA diminishes for $r_0$ = 20 bohr, the $n$ = 1 channel
is lower by 3-4 orders of magnitude and the $n$ = 2 channel starts to miss some
of the structures. Even inclusion of the first-order couplings starts to show
small but visible differences when compared to the exact results.

Deterioration of the BOA results at large electronic distances is a general
knowledge in the field of molecular physics that deals with the bound states.
In case of continuum states the \textit{R}-matrix poles become denser for larger
electronic box radii $r_0$ as shown in Fig.~\ref{fig-Ep}. On the other,
Fig.~\ref{fig-Vc1} demonstrates that the first-order nonadiabatic coupling elements
$\braket{\psi_k'}{\psi_{k'}}_r$ of Eq.~(\ref{eq-Vc1}) do not follow this behavior as
their magnitude is relatively insensitive to the $r_0$. Therefore it is clear that for
increasing electronic box size $r_0$,
the strength of the coupling terms $V^{(1)}_{k k'}(X)$ on the r.h.s.
of Eq.~(\ref{eq-nuc-coupl}) will become comparable with the spacing 
of the adiabatic curves of \textit{R}-matrix poles $E_k(R)$ in Fig.~\ref{fig-Ep}.
At this moment the Born-Oppenheimer solutions inside the electronic box confined
by the $r_0$ will cease to be valid. Depending on the desired accuracy, the present
model indicates that this may happen already for $r_0 < 12$ bohr. 

\begin{figure}[bht]
\includegraphics[width=0.46\textwidth]{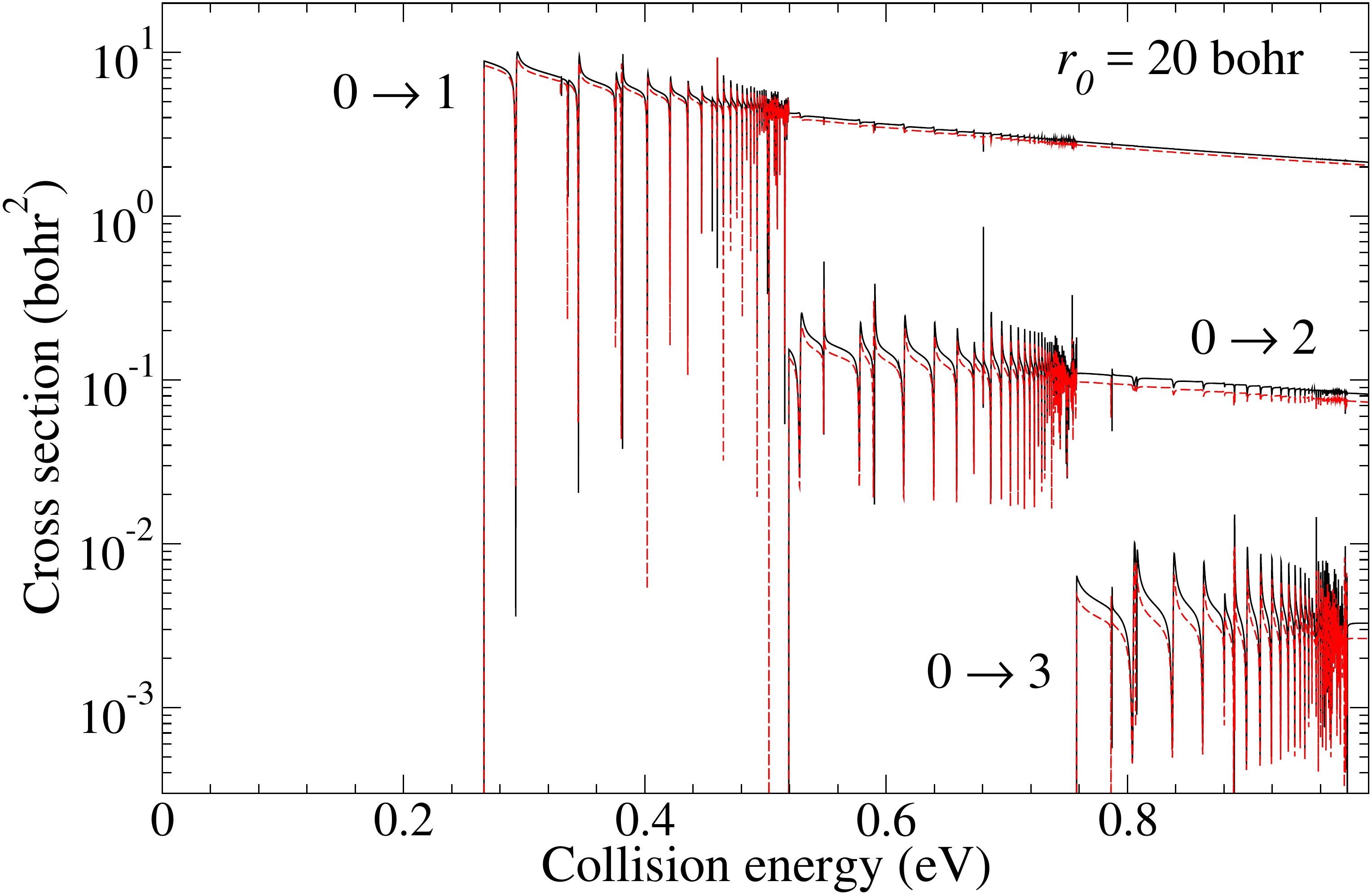}
\caption{\label{fig-vib}
Vibrationally inelastic cross sections. Full black curves represent the
exact results, while the red dashed show calculations from the Born-Oppenheimer
\textit{R}-matrix and $r_0$ = 20 bohr.
}
\end{figure}

In order to complete the present analysis we have also carried out a similar study
for the second, non-reactive channel that is an inseparable part of the calculations.
Fig.~\ref{fig-vib} shows a comparison between the exact electron-impact vibrational
excitation cross section and those obtained from the BOA \textit{R}-matrix. 
The data are displayed for the largest \textit{R}-matrix radius $r_0$ = 20 bohr.
Considering the failure of the Born-Oppenheimer approximation in case
of the DR channel displayed
in Fig.~\ref{fig-boa_r20}, one observes that the BOA has much weaker impact on the
vibrationally inelastic process. Moreover, we do not present the BOA vibrationally
inelastic cross sections for smaller \textit{R}-matrix radii $r_0$ = 6, 12 bohr,
because they are practically indistinguishable from the exact results.

\section{\label{sec-con}Conclusions}

The Born-Oppenheimer approximation is a cornerstone of all the 
ab-initio techniques
employed in the practical description of elastic and inelastic collisions
of electrons with molecules or molecular cations. These techniques involve
either the BOA \textit{R}-matrix method of \citet{Schneider_BoA_Rmat_JPB_1979}
or various forms of energy-dependent or energy-independent frame transformation
methods. It is important to emphasize that in the present study the BOA is 
considered only at short-range electronic distances $r_0 \leq$ 20 bohr while
the long-range parts of the involved Rydberg states are treated analytically.

In order to assess the validity of the short-range 
Born-Oppenheimer approximation beyond
the experimental accuracy we studied a 2D realistic model describing collisions
of electrons with H$_2^+$ in the singlet ungerade symmetry. We proposed
the 2D \textit{R}-matrix method to solve this model exactly (within the numerical
accuracy) for the dissociative recombination and the vibrational excitation channels.
The procedure of the exact solution is separated into two steps. 

In the first step all
the coupling electron-nuclear interactions are involved in determination of the
2D \textit{R}-matrix on the surface encompassing the region of these interactions.
In case this surface is too small to fit the electronic channels in the dissociative
process (quite common for the target cations), we also developed a technique to recompute
losslessly the \textit{R}-matrix on a surface of a larger 2D box.
Since the wave function determined at small distances also contains contribution from
closed channels, these contributions are eliminated simultaneously on the electronic
and nuclear surfaces. The elimination of the closed channel represents the full content
of the second step.

The exact results then serve as a benchmark for calculations in which the wave functions
inside the 2D box are represented by the Born-Oppenheimer products. We demonstrate that
for the DR channel the BOA starts to visibly break somewhere between 6-12 bohr of
the electronic \textit{R}-matrix radius $r_0$. Such a narrow validity of the BOA is 
very impractical because for most of the ab-initio calculations we expect $r_0$ $>$ 15
bohr for all the internuclear distances $R$ involved. We also show that the
first-order nonadiabatic coupling terms correct the inaccuracy of the BOA up to the highest
studied $r_0$ = 20 bohr. However, such couplings are very difficult to implement in the
present ab-initio \textit{R}-matrix codes
\cite{Morgan_Chen_Rm_1997}. The need for the nonadiabatic coupling terms was already
recognized by \citet{Sarpal_TM_DR_HeH_JPB_1994}, who used the diabatic representation
to numerically estimate the nonadiabatic couplings.

We also demonstrate that the situation is much better in case of the vibrational excitation
channels. The discrepancies between the exact and BOA results, found for the largest
\textit{R}-matrix radius $r_0$ = 20 bohr, are less than 10\% for the dominant
0 $\rightarrow$ 1 transition.

\appendix

\section{\label{sec-app}\textit{R}-matrix propagation in two dimensions}

\begin{figure}[htb]
\includegraphics[width=0.45\textwidth]{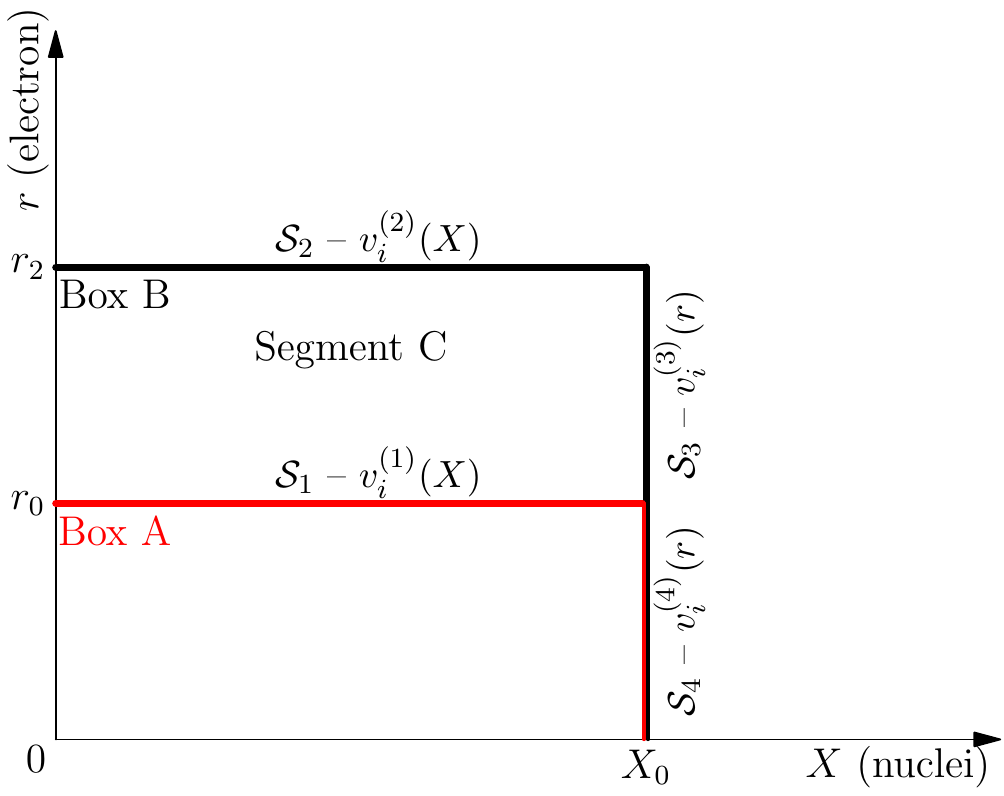}
\caption{\label{fig-prop2d}
Propagation of the 2D \textit{R}-matrix from Box A to Box B. The two boxes share
the surface $\mathcal{S}_4$. Functions $v^{(\alpha)}$ denote a complete set of orthonormal
functions defined on the respective surface $\mathcal{S}_{\alpha}$.
}
\end{figure}

Aim of this appendix is to derive a technique that allows to recompute an
\textit{R}-matrix defined on the surface of the Box A (see Fig.~\ref{fig-prop2d})
to the surface of the Box B. As can be seen in Fig.~\ref{fig-prop2d}, the Box A
is surrounded by surfaces $\mathcal{S}_1$ and $\mathcal{S}_4$, while the complete set
of the orthonormal functions on the surface encompassing the Box B is formed from
subsets defined on surfaces $\mathcal{S}_2$, $\mathcal{S}_3$, and $\mathcal{S}_4$.

The derivation here is a straightforward generalization of the one dimensional
\textit{R}-matrix propagator by 
\citet{Baluja_BM_CPC_1982}. 
A technique similar to the one presented here was also implemented by
\citet{Scott_2DRM_CPC_2009} for a two-dimensional \textit{R}-matrix propagation.
Their procedure is tailored for two indistinguishable particles (electrons) while
the present model deals with one electronic and one nuclear degree of freedom.

In the first step we diagonalize the symmetrized
Hamiltonian in the Segment C formed by the difference between the Box B 
and the Box A. Before the diagonalization the total Hamiltonian 
(\ref{eq-bloch2}) needs to be symmetrized by the Bloch operator
\begin{equation}
L = \frac{1}{2}
\left[
\delta(X\!-\!X_0)\frac{\partial}{\partial X} \!+\!
\delta(r\!-\!r_2)\frac{\partial}{\partial r} \!-\!
\delta(r\!-\!r_0)\frac{\partial}{\partial r}
\right],
\end{equation}
which ensures that $H+L$ is Hermitian on the Segment C for functions with
arbitrary boundary conditions on surfaces $\mathcal{S}_1$, 
$\mathcal{S}_2$, and $\mathcal{S}_3$.

After diagonalization of the $H+L$ operator in the Segment C
\begin{equation}
\label{eq-app-Hup}
(H + L) \ket{u_p} = E_p \ket{u_p} \;,
\end{equation}
the solution of the Schr\"{o}dinger equation (\ref{eq-Schrodinger-2DX}) can
be expanded in the Segment C as
\begin{equation}
\label{eq-app-u}
\ket{u} = \sum_p \ket{u_p}\frac{\bra{u_p} L \ket{u}}{E_p - E}.
\end{equation}
Furthermore, the solution $\ket{u}$ and its surface derivative can be evaluated
on the surfaces $\mathcal{S}_1$, $\mathcal{S}_2$, $\mathcal{S}_3$ surrounding
the Segment C and then projected onto respective complete set of surface functions 
$v^{(\alpha)}_i$ ($\alpha=1,2,3$) as
\begin{equation}
u^{\alpha}_i = (v^{(\alpha)}_i|u), \quad\quad u'^{\alpha}_i = (v^{(\alpha)}_i|u'),
\end{equation}
where $u'$ is a normal derivative on the respective surface and $(.|.)$
denotes scalar product over the surface coordinate.
Surface projections of Eq.~(\ref{eq-app-u}) can be now written in the following
compact vector equation:
\begin{equation}
\label{eq-app-u123}
u^{\alpha} = - \underline{\mathcal{R}}^{\alpha 1}.u'^1 + 
\underline{\mathcal{R}}^{\alpha 2}.u'^2 + \underline{\mathcal{R}}^{\alpha 3}.u'^3,
\quad\alpha=1,2,3.
\end{equation}
Matrix elements of the six independent matrices 
$\underline{\mathcal{R}}^{\alpha\beta}$ are
\begin{equation}
\label{eq-app-Rab}
\mathcal{R}^{\alpha\beta}_{ij} = \frac{1}{2} \sum_p 
\frac{( v^{(\alpha)}_i | u_p )( u_p | v^{(\beta)}_j)}{E_p - E},
\quad\alpha=1,2,3.
\end{equation}

\begin{figure}[!ht]
\includegraphics[width=0.48\textwidth]{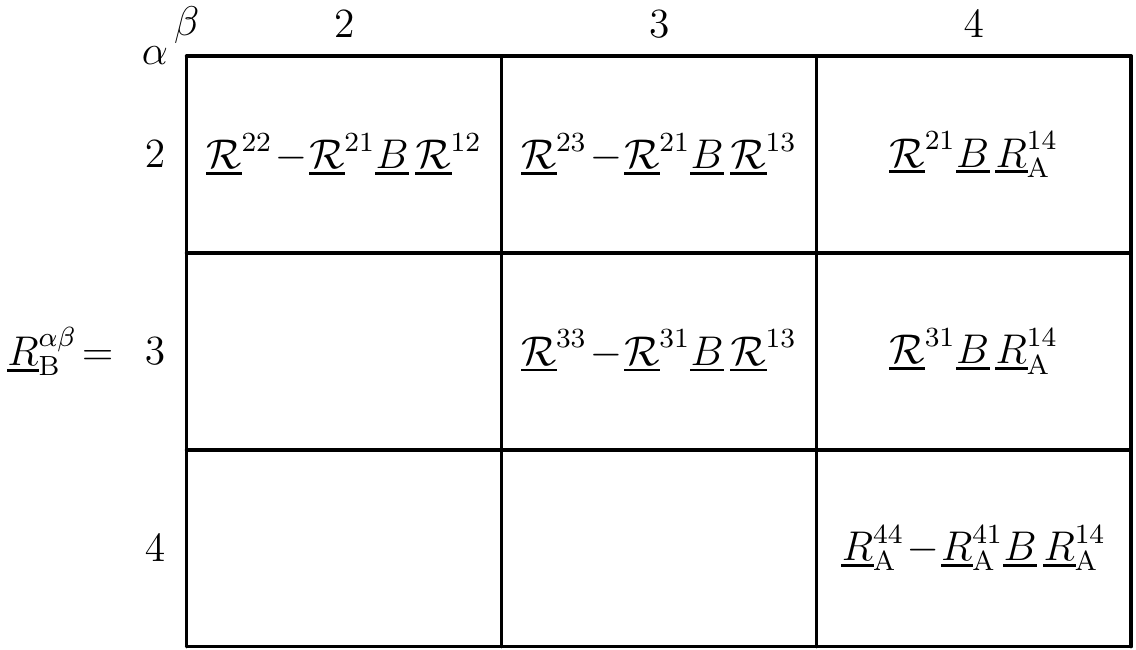}
\caption{\label{fig-Rb}
Matrix elements of the Block B \textit{R}-matrix constructed on surfaces 
$\mathcal{S}_2$, $\mathcal{S}_3$, and $\mathcal{S}_4$.
}
\end{figure}

An input for the propagation procedure presented here is the \textit{R}-matrix
$\underline{R}_{\mathrm{A}}$ for the Box A coupling the values and the normal derivatives 
on surfaces $\mathcal{S}_1$ and $\mathcal{S}_4$ as
\begin{equation}
\label{eq-app-u14}
u^{\alpha} = \underline{R}_{\mathrm{A}}^{\alpha 1}.u'^1 + 
\underline{R}_{\mathrm{A}}^{\alpha 4}.u'^4, 
\quad\alpha=1,4.
\end{equation}
The \textit{R}-matrix $\underline{R}_{\mathrm{B}}$ for the Box B will couple the
surfaces $\mathcal{S}_2$, $\mathcal{S}_3$, and $\mathcal{S}_4$ via
\begin{equation}
\label{eq-app-u234}
u^{\alpha} = \underline{R}_{\mathrm{B}}^{\alpha 2}.u'^2 + 
\underline{R}_{\mathrm{B}}^{\alpha 3}.u'^3 + 
\underline{R}_{\mathrm{B}}^{\alpha 4}.u'^4, 
\quad\alpha = 2,3,4.
\end{equation}
Combining Eqs.~(\ref{eq-app-u123}), (\ref{eq-app-u14}), and (\ref{eq-app-u234})
we arrive to the matrix elements of $\underline{R}_{\mathrm{B}}^{\alpha\beta}$
shown in Fig.~\ref{fig-Rb}, with the matrix $\underline{B}$ defined on the
surface $\mathcal{S}_1$ as
\begin{equation}
\underline{B} = \left( \underline{\mathcal{R}}^{11} + 
\underline{R}_{\mathrm{A}}^{11} \right)^{-1}.
\end{equation}
From the definition of matrices $\mathcal{R}^{\alpha\beta}_{ij}$ 
(\ref{eq-app-Rab}) it is clear that the result of the 2D propagation,
the matrix $\underline{R}_{\mathrm{B}}$ is Hermitian
on the surface surrounding the Box B, provided the 
matrix $\underline{R}_{\mathrm{A}}$ was Hermitian in the first place.

We conclude this section with two remarks of a technical nature:
\begin{itemize}
\item
While we have not assumed any particular form of the Hamiltonian 
(\ref{eq-app-Hup}) diagonalized in the Segment C, in most of the
practical applications as in the present study, the Hamiltonian
$H$ becomes separable in the nuclear and electronic coordinates. This
trivializes the formal 2D diagonalization in (\ref{eq-app-Hup}) to two
one dimensional diagonalizations.
In such a case the cost of all the operations needed to recompute the
2D \textit{R}-matrix from the Box A to the Box B is $\sim N^3$, where
$N$ is the size of the one-dimensional basis. It can be compared to the
cost $\sim N^6$ of the 2D diagonalization inside the Box A.
\item
The final form of the $\underline{R}_{\mathrm{B}}$-matrix displayed
in Fig.~\ref{fig-Rb} is expressed in a complete set of orthonormal channels
on the surface of the Box B. However, these channels do not represent the
physical channels into which the nuclei dissociate. 
The
matrix $\underline{R}_{\mathrm{B}}$ needs to be transformed into the
proper physical channels (\ref{eq-chan1}), (\ref{eq-chan2}) 
defined on the surface surrounding the Box B as
\begin{equation}
\left( R_{\mathrm{B}}\right)_{ij} = 
\sum_{\alpha,\beta=2}^4 (i|v^{(\alpha)}) .
\underline{R}_{\mathrm{B}}^{\alpha\beta} . (v^{(\beta)}|j)\;.
\end{equation}
\end{itemize}

\begin{acknowledgments}
R.\v{C}. and D.H. conducted this work with support of the Grant Agency of Czech Republic
(Grant No. GACR 18-02098S).
The contributions of C.H.G. were supported in part by
the U.S. Department of Energy, Office of Science, under Award No. DE-SC0010545.
\end{acknowledgments}

\bibliographystyle{apsrev}
\bibliography{DR}

\end{document}